\documentclass[letterpaper, 10 pt, conference]{ieeeconf}  % Comment this line out if you need a4paper

\IEEEoverridecommandlockouts                              % This command is only needed if 
\usepackage{amsmath}
\usepackage{amsfonts}
\usepackage{amssymb}
\usepackage{gensymb}
\usepackage{bm}
\usepackage{algorithm}
\usepackage{algpseudocode}
\usepackage{hyperref}
\usepackage{url}
\usepackage{epsfig}
\usepackage{graphicx}
\graphicspath{{figures}}

\def\rmd{\mathrm{d}}
\def\cN{\mathcal{N}}
\def\-{\text{-}}
\def\+{\text{+}}

\newcommand\given[1][]{\:#1\vert\:}

\usepackage[]{todonotes}

\title{\LARGE \bf
Bayesian grey-box identification\\ of nonlinear convection effects in heat transfer dynamics
}

\author{Wouter M. Kouw, Caspar Gruijthuijsen, Lennart Blanken, Enzo Evers, Timothy Rogers% <-this % stops a space
\thanks{Kouw and Blanken are with TU Eindhoven. Gruijthuijsen, Evers and Blanken are with Sioux Technologies B.V., Eindhoven, the Netherlands, and are supported by the Intelligent Motion Control consortium project. Rogers is with Sheffield University, Sheffield, United Kingdom, and is supported by EPSRC u/w/002140/1. Corresponding email: {\tt\small w.m.kouw@tue.nl}% <-this % stops a space
}
}

\begin{document}

\maketitle
\thispagestyle{empty}
\pagestyle{empty}

%%%%%%%%%%%%%%%%%%%%%%%%%%%%%%%%%%%%%%%%%%%%%%%%%%%%%%%%%%%%%%%%%%%%%%%%%%%%%%%%
\begin{abstract}
We propose a computational procedure for identifying convection in heat transfer dynamics. The procedure is based on a Gaussian process latent force model, consisting of a white-box component (i.e., known physics) for the conduction and linear convection effects and a Gaussian process that acts as a black-box component for the nonlinear convection effects. States are inferred through Bayesian smoothing and we obtain approximate posterior distributions for the kernel covariance function's hyperparameters using Laplace's method. The nonlinear convection function is recovered from the Gaussian process states using a Bayesian regression model. We validate the procedure by simulation error using the identified nonlinear convection function, on both data from a simulated system and measurements from a physical assembly.
\end{abstract}

%%%%%%%%%%%%%%%%%%%%%%%%%%%%%%%%%%%%%%%%%%%%%%%%%%%%%%%%%%%%%%%%%%%%%%%%%%%%%%%%
\section{Introduction}
Motion control systems are becoming ever more demanding in terms of throughput and accuracy, leading to increased attention to thermally induced deformations that cause slow drifts during positioning \cite{evers2019identifying,evers2020fast}.
The challenge in modelling these deformations lies in capturing the effects of convection on heat transfer. A typical approach is to describe the airflow surrounding the motion control system in detail, but this requires extensive expert knowledge and considerable computational resources. Recently, hybrid model- (white-box) and data-driven (black-box) approaches have been proposed to approximate the effects of the airflow, such as physics-informed neural networks \cite{cai2021physics}. We propose such a hybrid model- / data-driven method, i.e., a grey-box, where the effects of conduction, linear convection and heat input are expressed explicitly and the nonlinear effects of convection are captured by a regression model.

Our method is based on Gaussian process latent force models (GPLFM), originally developed to estimate unmeasured forces in mechanical systems \cite{alvarez2009latent,alvarez2013linear,sarkka2018gaussian,rogers2020bayesian}. For example, one could identify the strength of the restoring force in a nonlinear oscillator \cite{rogers2022latent}. But GPLFMs have been used to estimate other latent functions over time, such as identifying thermal dynamics in structures \cite{ghosh2015modeling}. We extend this work by tackling the effects of convection in heat transfer dynamics. GPLFMs are based on the conversion of temporal Gaussian processes (GP) to stochastic differential equations, allowing them to be incorporated into state-space models \cite{hartikainen2010kalman,hartikainen2012sequential}. This step enables online estimation of latent functions and simplifies the calculation of uncertainty estimates, two features not shared with physics-informed neural networks.
%
% Convection is position-dependent and will require a function estimator for each component in a lumped-element model. This means the number of states in the state-space model doubles, but we shall see that the linear nature of inference ensures the identification procedure remains efficient. 
% There is natural convection which is caused by the machine's heat producing an airflow of its own, and there is active convection, which is caused by external sources such as cooling fans. Separating natural from active convection requires an external signal to indicate when the active element is on.

% \wouter{todo: \cite{frigola2014variational}, \cite{daunizeau2009variational}}
% \tim{Definitely the work from Frigola is very nice and relevant but the computational load is very big, even more for their SMC version}

Our contribution consists of the application of a GPLFM to identify convection in heat transfer dynamics, quantification of uncertainty over the identified nonlinear convection function and validation of the proposed procedure on both simulated data and physical measurements.

\section{Model specification}
We briefly review heat transfer dynamics and GPLFMs. We then demonstrate how the two may be combined. 
% Section \ref{sec:inference} will show how to estimate temperature and convection effects.

\subsection{Heat transfer dynamics}
Consider a lumped-element model of a thermal system with $D$ components. Let $T_i(t) \in \mathbb{R}$ be the temperatures at time $t$ in each of the components, $T_a(t) \in \mathbb{R}$ be the ambient temperature and $u_i(t) \in \mathbb{R}^{+}$ be heat input. The evolution of these temperatures is assumed to be governed predominantly by conduction, convection and heat input:
\begin{align}
    M \dot{T} = \underbrace{K T}_{\textrm{conduction}} + \underbrace{h \big( T, T_a \big)}_{\textrm{convection}} + \underbrace{u}_{\textrm{input}} \, .
\end{align}
The dependence on $t$ will often be omitted for the sake of brevity in the remainder of the article.
The diagonal matrix $M$ represents the mass $m_i \in \mathbb{R}^{+}$ of each component multiplied with the specific heat capacity $\mathrm{c}_{p,i} \in \mathbb{R}^{+}$ of the component's material.
% \begin{align}
%     M = \begin{bmatrix} m_1 c_{p,1} & 0 & 0 \\ 0 & \ddots & 0 \\ 0 & 0 & m_N c_{p,N} \end{bmatrix} 
% \end{align}
The conductance matrix $K$ describes how heat is shared between components, i.e., $k_{i,j} \in \mathbb{R}^{+}$ indicates how much heat is conducted from component $i$ to component $j$. 

Convection is the loss of heat due to exchange with the medium surrounding the mechanical system. It can be split into a linear and a nonlinear term:
\begin{align}
    \underbrace{h \big( T_i, T_a \big)}_{\textrm{total cooling}} = \underbrace{h_a a_i \big(T_a  - T_i \big) }_{\textrm{linear convection}} + \underbrace{r \big(T_i, T_a \big)}_{\textrm{nonlinear convection}} \label{eq:convection} \, .
\end{align} 
The linear convection term describes the loss of heat proportional to the surface area $a_i$ times the difference between the temperature of the material $T_i$ and the ambient temperature $T_a$. We assume uniform cooling over the surface with a heat transfer coefficient $h_a$. 
The remainder is a nonlinear function of the temperature of the material and the ambient temperature. We call this $r(T_i,T_a)$ the nonlinear convection function. Proper physics-driven modelling would employ computational fluid dynamics and include explicit dependencies on the airflow of the surrounding environment. 
% A white-box model would incorporate these variables, but here we treat the output of this function as a standard unknown variable.
Here, we do not model these terms but will instead capture the effect of the nonlinear convection function, i.e., the output of $r(T,T_a)$, using a black-box function approximator (c.f. Section \ref{sec:assm}).

The ambient temperature is assumed to be measured and will be treated as an input. If we incorporate the linear convection term into the governing equations, they become:
\begin{align} \label{eq:heatdynamics}
    \dot{T} = M^{-1} F T + M^{-1}r\big(T_a, T \big) + M^{-1} G \bar{u} \, ,
\end{align}
where $\bar{u} = [ T_a \ u_1 \ \dots \ u_D]^{\intercal}$ and
\begin{align}
     F = K - \begin{bmatrix} h_a a_1  & & \\ & \! \ddots \! & \\ & &  h_a a_D  \end{bmatrix} , \, G= \begin{bmatrix} \begin{matrix} h_a a_1  \\ \vdots \\ h_a a_D  \end{matrix}  & I \ \end{bmatrix} . 
\end{align}
%
% We shall capture the effect of the nonlinear convection term using a Gaussian process.

\subsection{Temporal Gaussian Processes}
Temporal Gaussian processes describe distributions over functions of time \cite{rasmussen2006gaussian}. We shall use these to estimate the state $\rho(t) = r(T(t),T_a(t))$ over time, and later fit a regression model from $T(t)$ and $T_a(t)$ to $\rho(t)$ (see Section \ref{sec:static-nonlin}) \cite{alvarez2013linear,sarkka2018gaussian}.
To do so, we must construct a dynamical form of a temporal Gaussian process. Consider a GP prior distribution over functions $\rho(t)$
\begin{align}
    p\big(\rho(t) ; \psi \big) = \mathcal{GP}\big(\rho(t) \given 0, \kappa_{\psi}(t,t')\big) \, ,
\end{align} 
with kernel covariance function $\kappa$ with hyperparameters $\psi$.
We have chosen a zero-mean prior distribution because we have no reason to believe our function of interest has a systematic offset. 
% We will consider a Whittle-Matern kernel defined as:
% \begin{align}
% \kappa(\tau) = \sigma^2 \frac{2^{1-\nu}}{\Gamma(\nu)} \left(\frac{\sqrt{2\nu}\tau}{l} \right)^\nu K_\nu\left(\frac{\sqrt{2\nu}\tau}{l} \right) \, ,
% \end{align}
% where $\sigma^2$ is a scale hyperparameter, $l$ a characteristic length-scale, $\nu$ the smoothness hyperparameter, and $K_\nu(.)$ is a modified Bessel function of the second kind. 
For the kernel covariance function, we select the lowest order of the Whittle-Mat{\'e}rn class, namely the exponential covariance function
\begin{align}
    \kappa_{\psi}(t,t') = \gamma^2 \exp \Big(- \frac{\sqrt{3}}{l} \, | \, t - t' \, | \Big) \, ,
\end{align}
% \wouter{todo: add motivation for exponential as kernel cov - Tim: I have added a sentence here}
with scale hyperparameters $\psi = (\gamma, l)$ \cite{rasmussen2006gaussian}. 
Note that this kernel is stationary, i.e., only a function of $t - t'$.
The choice for an exponential covariance is based on two qualitative features: it is flexible (we make no strong smoothness assumptions) and it leads to a scalar dynamical systems (see also Sec.~\ref{sec:discussion}).
% The exponential covariance provides a flexible nonlinear kernel for stationary functions which are not expected to be smooth.
%
% After observing data, one would typically build a matrix filled with evaluations of the kernel covariance function and compute a posterior distribution over functions using Bayes' rule, see Rasmussen and Williams \cite{rasmussen2006gaussian} for details. The analytical posterior requires the inversion of the $N \times N$ matrix, which is notoriously expensive (naive implementations cost $O(N^3)$). But temporal Gaussian processes can be re-expressed as linear stochastic differential equations, thus allowing for inference in $O(N)$ \cite{hartikainen2010kalman}. Indeed, t

The exponential kernel covariance function is dual to the power spectral density \cite{hartikainen2010kalman}:
\begin{align} \label{eq:power-density}
   \mathcal{K}_{\psi}(\omega) \propto \frac{2\lambda \gamma^2}{(\lambda^2 + \omega^2)} \, , 
\end{align}
where $\lambda = \sqrt{3} / l$. Factorization of the denominator produces a transfer function that serves as the state transition in the stochastic differential equation (SDE) \cite{hartikainen2010kalman};
\begin{align} \label{eq:gp-SDE}
\dot{\rho}(t) = - \lambda \rho(t) + w(t) \, .
\end{align}
The white noise process $w(t)$ has a spectral density equal to the numerator of Eq.~\ref{eq:power-density}, $v_c = 2\lambda \gamma^2$. 
% The above SDE can be mapped to a state-space model with a prediction-correction inference procedure for its states (c.f. Section \ref{sec:inference}). Its predictions at time points $t_{*}$ are equivalent to the predictions by the original GP formulation \cite{hartikainen2010kalman} when considering the smoothing distribution.

\subsection{Augmented discrete-time model} \label{sec:assm}
In this section, we will augment the system of differential equations (Eq.~\ref{eq:heatdynamics}) with the SDE representation of the Gaussian process. 
Note that the nonlinear convection function $r(T, T_a)$ is a vector and that Eq.~\ref{eq:gp-SDE} describes a scalar function. We therefore pose an independent GP SDE $\rho(t)$ for every component:
\begin{align}
    r(T(t),T_a(t)) = \begin{bmatrix} r(T_1(t),T_a(t)) \\ \vdots \\ r(T_D(t),T_a(t)) \end{bmatrix} \approx \begin{bmatrix} \rho_1(t) \\ \vdots \\ \rho_D(t) \end{bmatrix} \, .
\end{align}
Using $\rho(t) = [\rho_1(t) \dots \rho_D(t)]^{\intercal}$, we can reformulate Eq.~\ref{eq:heatdynamics} as an augmented system:
\begin{align}
\begin{bmatrix} \dot{T} \\ \dot{\rho} \end{bmatrix}  =  \begin{bmatrix} M^{\-1} F & M^{\-1} \\ 0 & -\lambda I \end{bmatrix} \begin{bmatrix} T \\ \rho \end{bmatrix}  +  \begin{bmatrix} M^{\-1}G \\  0  \end{bmatrix} \bar{u}  +  \begin{bmatrix} 0 \\ 1 \end{bmatrix} w ,
\end{align}
where $\lambda$ and $\gamma$ are shared across all GP states.

We shall discretize the system using a regular sampling interval $\Delta t = t_k - t_{k-1}$ for all steps $k$. Let $x_k = [T_{1k} \dots T_{Dk} \ \rho_{1k} \dots \rho_{Dk}]^{\intercal}$. The system then becomes:
\begin{align}
x_{k} = A x_{k\-1} + B \bar{u}_k + w_k\, ,
\end{align}
with state transition and control matrix
\begin{align}
    A = \exp\big(\Delta t \begin{bmatrix} M^{\-1}F & M^{\-1} \\ 0 & -\lambda I \end{bmatrix} \big) \, , \ B = \Delta t \begin{bmatrix}  \, M^{\-1} G \\ 0 \end{bmatrix}  \, .
\end{align}
The discrete-time noise $w_k$ is zero-mean Gaussian distributed, with covariance matrix
\begin{align} \label{eq:noise-cov}
    Q = \int_0^{\Delta t} \exp(A t) \begin{bmatrix} 0 \\ I \end{bmatrix} (v_c I) \begin{bmatrix} 0 \\ I \end{bmatrix}^{\intercal} \exp(A t)^{\intercal} dt \, .
\end{align}
We approximate this integral using a first-order Taylor approximation of the matrix exponential: $\exp(A t) \approx I + At$. This provides an analytic expression that can be easily differentiated (important for Sec.~\ref{sec:hparams-estimation}).
The approximation \eqref{eq:noise-cov} evaluates to 
\begin{align}
    Q &\approx \begin{bmatrix} Q_{11} & Q_{12} \\ Q_{21} & Q_{22} \end{bmatrix} \, ,
\end{align}
with block matrices
\begin{subequations}
\begin{align}
    Q_{11} &= \frac{1}{3}\Delta t^3 v_c M^{-1}\\
    Q_{12} &= Q_{21} = (\frac{1}{2}\Delta t^2 - \frac{1}{3}\lambda \Delta t^3) v_c M^{-1} \\
    Q_{22} &= (\Delta t - \lambda \Delta t^2 + \frac{1}{3}\lambda^2 \Delta t^3) v_c I \, .
\end{align}
\end{subequations}
Note that both $A$ and $Q$ depend on the hyperparameters $\lambda$ and $\gamma$, and will henceforth be referred to as $A_\psi$ and $Q_\psi$.

\subsection{Probabilistic state-space model}
Our goal will be to infer the temperature and GP states, for which we required a probabilistic state-space model. If we integrate out the process noise instance $w_k$, then the distribution of the next state $x_{k+1}$ is Gaussian:
\begin{align}
    p(x_{k} \given x_{k\-1}, \bar{u}_k ; \psi) = \mathcal{N}(x_{k} \given A_{\psi} x_{k\-1} + B \bar{u}_k, Q_{\psi}) \, .
\end{align}
We assume to have noisy measurements of the temperatures:
\begin{align}
    p(y_k \given x_k) = \mathcal{N}( y_k \given C x_k, R) \, ,
\end{align}
where $C$ indicates which components are measured and $R$ is the measurement noise covariance matrix.

The prior distribution of the temperatures is assumed to be Gaussian distributed;
\begin{align}
    p(T_0) = \mathcal{N}\big(T_0 \given \hat{m}_0, \hat{S}_0 \big) \, .
\end{align}
For the SDE of the temporal Gaussian process to be stable, it must start from the steady-state solution of the process. 
% In other words, the time derivatives of the state mean $\bar{f}$ and variance $\bar{\bar{f}}$ must be zero;
% \begin{subequations}
%     \begin{align}
%     \frac{d \bar{f}_{\infty}}{dt} &= -\lambda \bar{f}_{\infty} = 0 \\
%     \frac{d \bar{\bar{f}}_{\infty}}{dt} &= -\lambda \bar{\bar{f}}_{\infty} - \bar{\bar{f}}_{\infty} \lambda^{\intercal} + 2\lambda \gamma^2 = 0 
%      \, .
% \end{align}
% \end{subequations}
% Solving these yield a stationary mean of $\bar{f}_{\infty} = 0$ and a stationary variance of $\bar{\bar{f}}_{\infty} = \gamma^2$.
Setting the time derivatives of the state distribution's parameters to $0$ yields a stationary mean of $0$ and, through Lyapunov's equation, a stationary variance of $\gamma^2$ \cite{sarkka2019applied}. 
The prior state distribution for the GP states thus becomes:
\begin{align}
    p(\rho_{i0}; \psi) = \mathcal{N}\big(\rho_{i0} \given 0, \gamma^2 \big) \, .
\end{align}
% \wouter{todo: add derivation of stationary covariance matrix}
% An important quantity is the stationary variance of the process, $v_\infty$, obtained by setting the time-derivative of the variance to $0$:
% \begin{align}
%     \frac{d P}{dt} = FP + PF^{\intercal} + L v_cL^{\intercal} = 0 \, .
% \end{align}
% The solution is $P_{\infty} = \gamma^2$.

Combining the heat transfer model and GP priors gives:
\begin{align}
    p(x_0; \psi) = \mathcal{N}\Big(\begin{bmatrix} T_0 \\ \rho_0 \end{bmatrix} \given \underbrace{\begin{bmatrix} \hat{m}_0 \\ 0 \end{bmatrix}}_{m_0}, \underbrace{\begin{bmatrix} \hat{S}_0 & 0 \\ 0 & \gamma^2 I\end{bmatrix}}_{S_0} \Big) \, .
\end{align}
The complete grey-box probabilistic model for a time-series of length $N$ is:
\begin{align}
    p(y_{1:N}, &\, x_{0:N} \given \bar{u}_{1:N} ; \psi)  = \\
    & \quad  p(x_0; \psi) \prod_{k=1}^N p(y_k \given x_k) p(x_k \given x_{k\-1}, \bar{u}_{k}; \psi) \, . \nonumber
\end{align}
We shall use this model to infer marginal posterior distributions over states $x_k$ and hyperparameters $\psi$.

\section{Inference} \label{sec:inference}
The inference procedure has two phases: firstly, states and hyperparameters are estimated, and secondly, nonlinear convection is estimated as a function of temperature.

\subsection{State estimation}
States are inferred using the Bayesian smoothing equations \cite{sarkka2023bayesian}. These start with a filtering step, moving from $k=1$ to $k=N$. Let $\mathcal{D}_k = \{y_i, \bar{u}_i\}_{i=1}^{k}$ be the input-output pairs up to time $k$. The prediction step is the marginalization of the Gaussian state transition over the previous Gaussian marginal state posterior:
\begin{align}
    p&(x_k \given \bar{u}_k, \mathcal{D}_{k\-1}; \hat{\psi}) \nonumber \\
    &= \! \int \! p(x_k \given x_{k\-1}, \bar{u}_{k}; \hat{\psi}) \, p(x_{k\-1} \given \mathcal{D}_{k\-1}; \hat{\psi}) \, \mathrm{d}x_{k\-1} \\
    % &= \! \int \! \cN(x_k | A_{\psi}x_{k\-1} \! + \! B\bar{u}_{1:k}, Q_{\psi}) \cN(x_{k\-1} | m_{k\-1}, S_{k\-1}) \rmd x_{k\-1} \\
    &= \cN\big(x_k \given \bar{m}_k, \bar{S}_{k} \big) \, ,
\end{align}
where the predictive mean and variance are
\begin{align} \label{eq:state_predict}
    \bar{m}_k = A_{\hat{\psi}} m_{k\-1} + B \bar{u}_k \, , \quad
    \bar{S}_k = A_{\hat{\psi}} S_{k\-1} A_{\hat{\psi}}^{\intercal} + Q_{\hat{\psi}} \, .
\end{align}    
Note that the kernel hyperparameters are fixed to a point estimate, $\hat{\psi}$ (see Sec.~\ref{sec:hparams-estimation}).
% \wouter{Variational approximation to get rid of dependence on $\psi$? Then Laplace's approximation becomes variational Laplace; more novelty.}

In the correction step, we apply Bayes' rule using the predicted marginal state as prior distribution:
\begin{align}
    \underbrace{p(x_k \given \mathcal{D}_k; \hat{\psi})}_{\text{posterior}} &= \frac{\overbrace{p(y_k \given x_k)}^{\text{likelihood}}}{\underbrace{p(y_k \given \bar{u}_{k}, \mathcal{D}_{k\-1})}_{\text{evidence}}} \, \underbrace{p(x_k \given \bar{u}_k, \mathcal{D}_{k\-1}; \hat{\psi})}_{\text{prior}} \, ,
\end{align}
where the evidence is
\begin{align}
    p(y_k | \bar{u}_{k}, \mathcal{D}_{k\-1}) \! = \! \int \! p(y_k | x_k) p(x_k | \bar{u}_{k}, \mathcal{D}_{k\-1}; \hat{\psi}) \rmd x_{k} \, .
\end{align}
For our Gaussian distributed likelihood and Gaussian distributed predicted state distribution, this yields a Gaussian state posterior $\cN(x_k \given \tilde{m}_k, \tilde{S}_k)$ with parameters \cite{sarkka2023bayesian}:
\begin{subequations} \label{eq:state_correct}
    \begin{align}
\tilde{m}_k &= \bar{m}_k + \bar{S}_k C^{\intercal}(C\bar{S}_k C^{\intercal} + R)^{-1}(y_k - C\bar{m}_k) \\
\tilde{S}_k &= \bar{S}_k - \bar{S}_k C^{\intercal}(C\bar{S}_k C^{\intercal} + R)^{-1} C \bar{S}_k^{\intercal} \, .
\end{align}
\end{subequations}
% These are the familiar Kalman filter equations.
%
Smoothing consists of correcting these state estimates based on future data \cite{sarkka2023bayesian}:
\begin{align}
    &p(x_k \given \mathcal{D}_N; \hat{\psi}) = p(x_k \given \mathcal{D}_k; \hat{\psi}) \\
    &\quad \cdot \int \frac{p(x_{k+1} \given x_k, \bar{u}_{k+1}; \hat{\psi}) p(x_{k+1} \given \mathcal{D}_N; \hat{\psi})}{p(x_{k+1} \given \mathcal{D}_k; \hat{\psi})} \mathrm{d}x_{k+1} \nonumber \, .
\end{align}
These corrections are executed by the following updates, running backwards from $k=N \dots 1$:%
\begin{subequations} \label{eq:state_smoothen}
    \begin{align}
    G_k &= \tilde{S}_k A_{\hat{\psi}}^{\intercal}\bar{S}_{k+1}^{-1} \\
    m_k &= \tilde{m}_k + G_k \big(m_{k+1} - \bar{m}_{k+1}  \big) \\
    S_k &= \tilde{S}_k + G_k \big( S_{k+1} - \bar{S}_{k+1} \big) G_k^{\intercal} \, ,
\end{align}
\end{subequations}
where $G_k$ represents the strength of the correction by future observations. 

To obtain state estimates, the runtime is $O( D^3 N)$ due to inversions of $D \times D$ covariance matrices. This algorithm therefore excels in situations where $D$ is small (i.e., few components in a lumped-element model) and $N$ is large (long time-series).

\subsection{Hyperparameter estimation} \label{sec:hparams-estimation}
Tuning GP kernel hyperparameters can be challenging when the landscape is multi-modal or contains regions of divergence. Maximum likelihood estimation in those cases may lead to poor solutions \cite{svensson2015marginalizing}. Here we propose a Laplace approximation of the posterior distribution, for two reasons: one, the use of prior distributions may enforce convergence, and two, the approximate posterior variance provides a quick method for assessing the quality of selected hyperparameters.  

We assume the hyperparameters are independent of each other, so $p(\psi) = p(\gamma) p(l)$. Since the length scales are strictly positive, we choose to employ Gamma distributed prior distributions:
\begin{align}
    p(\gamma) = \mathcal{G}(\gamma \given \alpha_{\gamma}, \beta_{\gamma}) \, , \quad p(l) =  \mathcal{G}(l \given \alpha_{l}, \beta_{l}) \, ,    
\end{align}
We then form a Gaussian approximation of the posterior distribution whose mean $\hat{\psi}$ is the maximum a posteriori and whose precision matrix $\Lambda$ is based on the curvature at the maximum \cite{murphy2012machine}:
\begin{align} \label{eq:hparams_laplace}
    \hat{\psi} &= \underset{\psi \in \Psi}{\arg \max} \ \ln p(y_{1:N} \given \bar{u}_{1:N} 
    ; \psi) p(\psi)  \\
    \Lambda_{ij} &= - \frac{\partial^2}{\partial \psi_i \, \partial \psi_j}  \ln  p(y_{1:N} \given \bar{u}_{1:N}; \psi) p(\psi)  \Big|_{\psi = \hat{\psi}} \, .
    \end{align}
Note that the Hessian only has to be computed once.

% \wouter{todo: prior parameter notation messy}
% The Gamma prior distributions enforce a probability of 0 for non-positive values. As such, the optimization problem described in Eq.~\ref{eq:hparams_laplace} is naturally constrained.

\subsection{Static nonlinearity estimation} \label{sec:static-nonlin}
State estimation will give us estimates of $T_k$ and $\rho_k$. But $\rho_k$ only represents the value of $r(T_i,T_a)$ at time $t_k$. To obtain a functional form for $r(\cdot, \cdot)$, we employ a regression model that takes as inputs the estimates of $T_{ik}$ (i.e., means $m_{ik}$ for $i=1, \dots D$) as well as the measurements of $T_{ak}$, and as outputs takes the estimates of $\rho_{ik}$ (i.e., the means $m_{jk}$ for $j = i+D$). We consider specifically a Bayesian polynomial regression model because low-order polynomials typically suffice to capture the relatively slow change in convection over temperatures and because we aim to quantify uncertainty on the function estimate. Let $\phi : \mathbb{R}^2 \rightarrow \mathbb{R}^{D_\phi}$ be a polynomial basis expansion function, mapping a cell and ambient temperature to a $D_\phi$-dimensional space, and $\theta_i \in \mathbb{R}^{D_\phi}$ be regression coefficients. We consider a likelihood of the form,
\begin{align}
    p(m_{jk} | m_{ik}, T_{ak}, \theta_i) \! = \! \mathcal{N}(m_{jk} \given \theta_i^{\intercal} \phi(m_{ik}, T_{ak}), \sigma_j^2) \, ,
\end{align}
where $\sigma_j^2 = \frac{1}{N} \sum_{k=1}^N S_{jjk}$ is the average variance of the GP estimated states. Our prior distribution on the regression weights $\theta_i$ is:
\begin{align}
    p(\theta_i) = \mathcal{N}(\theta_i \given \mu_{i0}, \Sigma_{i0}) \, .
\end{align}
This Gaussian prior distribution is conjugate and we may thus obtain a posterior distribution exactly \cite{murphy2012machine},
\begin{align}
    p(\theta_{i} \given \mathcal{D}_N) &= \frac{p(\theta_i) \prod_{k=1}^N p(m_{jk} \given m_{ik}, T_{ak}, \theta_i)}{\int p(\theta_i) \prod_{k=1}^N p(m_{jk} \given m_{ik}, T_{ak}, \theta_i) \mathrm{d}\theta_i} \\
    &= \mathcal{N}(\theta_i \given \mu_{i}, \Sigma_{i}) \, ,
\end{align}
where, using $\phi_k = \phi(m_{ik}, T_{ak})$, the parameters are: 
\begin{align} \label{eq:post-regweights}
    \Sigma_{i} &= \big(\Sigma_{i0}^{-1} \! + \! \sigma_j^{-2} \sum_{k=1}^{N} \phi_k \phi_k^{\intercal} \big)^{-1} \\
    \mu_{i} &= \Sigma_i \big( \sigma_j^{-2} \sum_{k=1}^{N} \phi_k m_{jk} \! + \! \Sigma_{i0}^{-1} \mu_{i0} \big) \, .
\end{align}

Given a posterior distribution over the regression coefficients, we can derive a predictive distribution over new values of the nonlinear convection function. These are essentially predictions for the values of $r(T_i, T_a)$ but with a variance parameter indicating the amount of uncertainty originating from the estimated values of $\rho_{i}$ and $\theta_i$ \cite{murphy2012machine}. Let $\phi_{*} = \phi(m_{i*}, T_{a*})$. Then, the posterior predictive is:
\begin{align}
    p(m_{j*} \given  &m_{i*}, T_{a*}, \mathcal{D}_{N}) \nonumber \\
    &= \int p(m_{j*} \given m_{i*}, T_{a*}, \theta_i) p(\theta_{i} \given \mathcal{D}_N) \mathrm{d} \theta_i  \\
    &= \mathcal{N}(m_{j*} \given \mu_{i}^{\intercal} \phi_{*}, \, \phi_{*}^{\intercal} \Sigma_{i} \phi_{*} + \sigma_j^2 \big) \, .\label{eq:post-preds}
\end{align}
% Note that in order to obtain the posterior variance for the convection function, one is required to supply a variance of the temperature estimate, $S_{ii*}$. 

For ease of reference, we refer to the mean of the posterior predictive as: 
\begin{align} \label{eq:pp_mean}
    \hat{r}(T_{i*}, T_{a*}) = \mu_i^{\intercal} \phi(T_{i*}, T_{a*})
\end{align}
for an arbitrary numerical temperature $T_{i*}$.

% \begin{algorithm}
% \caption{Infer nonlinear convection function}\label{alg:inf}
% \begin{algorithmic}
% \Require $y_{1:N}, \bar{u}_{1:N}, K,M,S,\theta, m_0, S_0$
% \For{$k = 1, \dots, N$}
%     \State Compute state estimates $\tilde{m}_k, \tilde{S}_k$ by \eqref{eq:state_predict}, \eqref{eq:state_correct}
% \EndFor
% \For{$k = N, \dots, 1$}
%     \State Compute state estimates $m_k, S_k$ by \eqref{eq:state_smoothen}
% \EndFor
% \State Compute kernel hyperparameters $\hat{\psi}, \Lambda_{\psi}$ by \eqref{eq:hparams_laplace}.
% % \State Calculate $z$ according to \eqref{eq:}
% \State Compute regression coefficients' parameters $\bmu, \bSigma$ \eqref{eq:post-regweights}
% \State Return posterior predictive distribution \eqref{eq:post-preds}
% \end{algorithmic}
% \end{algorithm}

\section{Experiments}
\begin{figure}[htb]
    \centering
    \includegraphics[width=.95\columnwidth]{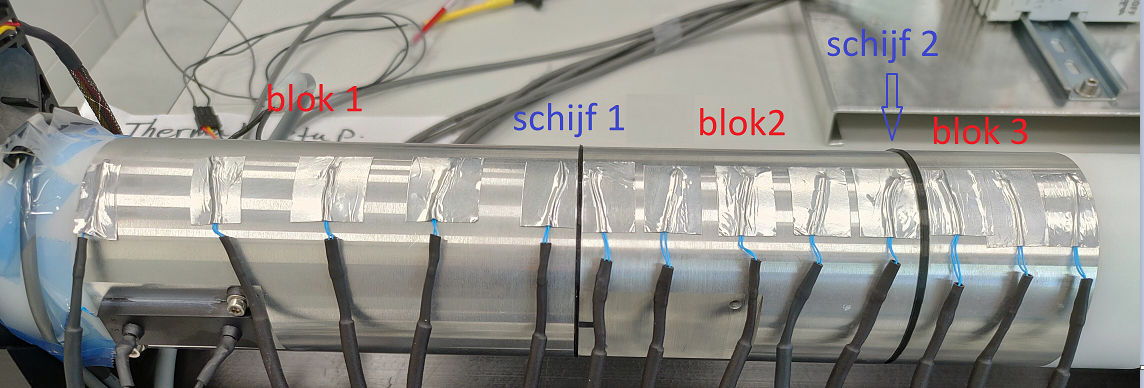} \\ \vspace{5pt}
    \includegraphics[width=.48\columnwidth]{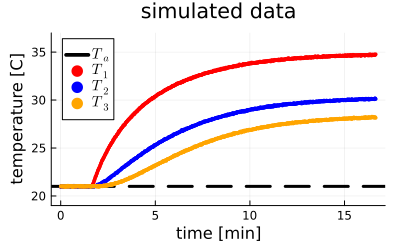}
    \includegraphics[width=.48\columnwidth]{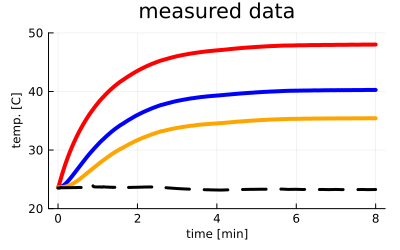} 
    \caption{(Top) Photo of heated rod demonstrator, with 3 blocks, 2 insulation discs, 13 temperature sensors and 1 heater. (Bottom left) Data simulated according to Eq. \ref{eq:heatdynamics} with parameters described in Sec.~\ref{sec:simdata}. (Bottom right) Data measured from demonstrator with parameters described in Sec.~\ref{sec:measureddata}.}
    \label{fig:experiment-setup}
    \vspace{-5pt}
\end{figure}

We perform experiments\footnote{Code: \url{github.com/biaslab/CCTA2024-BIDconvection}} on a heated rod demonstrator involving 3 aluminum blocks separated by insulating discs (see Figure \ref{fig:experiment-setup} top). The first experiment involves a simulation of the system and the second experiment involves measurements from the physical device. Only the first block receives heat input. The conductance matrix is of the form:
\begin{align}
    K = \begin{bmatrix} 
        -k_{12} & k_{12} & 0 \\ 
        k_{12} & -(k_{12} + k_{23}) & k_{23} \\
        0 & k_{23} & -k_{23} 
        \end{bmatrix} \, ,
\end{align}
whose $k_{ij}$ are assumed to be known or estimated separately using a maximum likelihood or MAP estimator. Further details are explained in the following two subsections.

\begin{figure*}[thb]
    \centering
    \includegraphics[width=.19\textwidth]{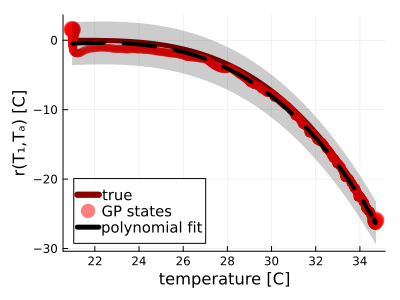}
    \includegraphics[width=.19\textwidth]{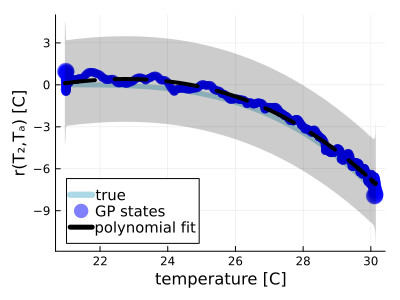}
    \includegraphics[width=.19\textwidth]{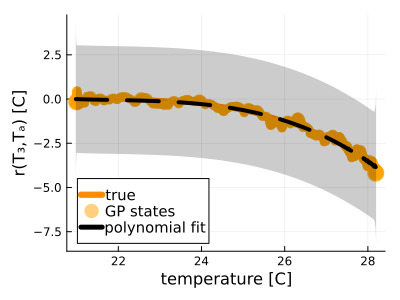}
    \includegraphics[width=.19\textwidth]{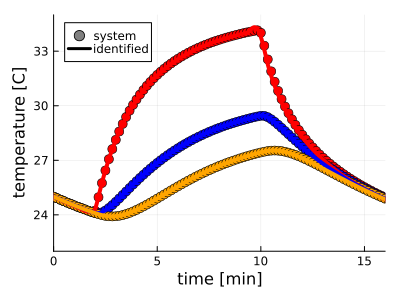} 
    \includegraphics[width=.19\textwidth]{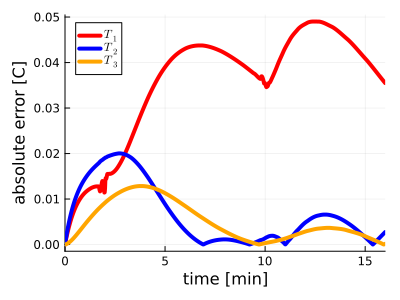}
    \\
    \includegraphics[width=.19\textwidth]{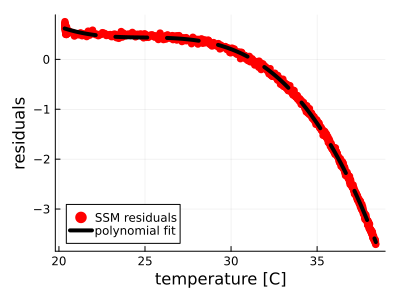}
    \includegraphics[width=.19\textwidth]{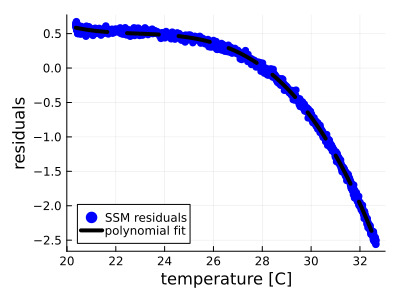}
    \includegraphics[width=.19\textwidth]{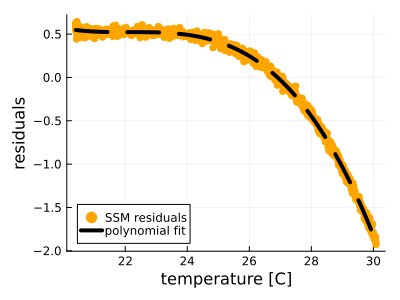}
    \includegraphics[width=.19\textwidth]{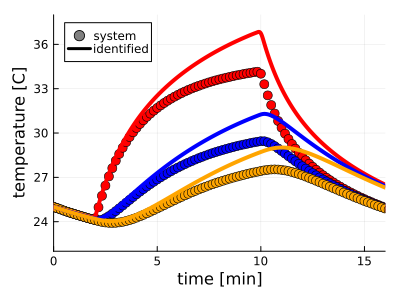}
    \includegraphics[width=.19\textwidth]{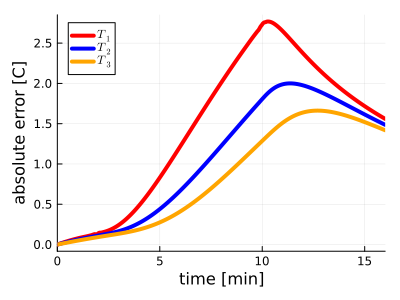}
    \caption{Experiment on data from simulated system. Top row: the first three subplots (left-to-right) show the value of the nonlinear convection function over visited temperature states, the GP estimates of this function and the posterior predictive distribution. The fourth subplot shows a forward simulation with the mean posterior predictive and the fifth subplot shows the absolute error (lower is better) between simulation with the true and identified convection function, i.e., $|$ states true - states identified $|$. Bottom row: first three subplots show standard polynomials fitted to residuals left by a state-space model without nonlinear convection terms. The fourth subplot shows a forward simulation using the regression function fit to the residuals, which deviates far from the forward simulation with the true convection function. The absolute error between the two is plotted in the fifth subfigure.}
    \label{fig:exp-simulated-data}
    \vspace{-5pt}
\end{figure*}

\subsection{Simulated data} \label{sec:simdata}
In the simulated system, each aluminum block is modeled as a uniform component. Conductance of the nylon pads is set to $k_{12} = k_{23} = 10 \, \mathrm{W}/(\mathrm{m} \mathrm{K}$) and the mass times specific heat capacity parameter is set to $m_i \mathrm{c}_{p,i} = 1000 \, \mathrm{J}/\mathrm{K}$ for all components. The ambient temperature is fixed to $21\,\degree \mathrm{C}$, the heat transfer coefficient $h_a$ is set to $2.0 \, \mathrm{W}/(\mathrm{m}^2 \mathrm{K})$ and the outer surface areas of all the blocks are set to $a_i = 1.0 \, \mathrm{m}^2$. The first block receives $100 \, \mathrm{W}$ of heating after $100$ seconds.
We used a simple surrogate for the nonlinear convection function: $r(T_a, T_i) = (T_a - T_i)^3 / 100$. Its main feature is that when the block temperature is much higher than the ambient temperature, convection will cause the block to cool rapidly. We simulated the temperatures forward in time for $1000$ steps of $\Delta t = 1$ seconds, using DifferentialEquations.jl \cite{rackauckas2017differentialequations}. Finally, we add zero-mean Gaussian distributed noise with a variance of $10^{-3}$ to the simulated states, to generate artificial measurements. The state prior distribution's parameters are $\bar{m}_0 = [21 \ 21 \ 21]$ and $\bar{S}_0 = I$. Bayesian smoothing was implemented with RxInfer.jl, which also produces the marginal likelihood in Eq.~\ref{eq:hparams_laplace} \cite{bagaev2023rxinfer}. Derivatives were computed with Optim.jl \cite{mogensen2018optim}. 

% \begin{figure}[ht]
%     % \centering
%     % \includegraphics[width=\columnwidth]{figures/nonlinear-convection.png} \\
%     \includegraphics[width=\columnwidth]{figures/exp-simulated-temperatures.png}
%     % \includegraphics[width=\columnwidth]{figures/expm-sim.png}
%     \includegraphics[width=\columnwidth]{figures/input-heat.png}
%     \caption{Experiment with simulated data. (Top) Temperature measurements of three blocks of a heated rod (red, blue, orange points for block 1, 2 and 3, respectively) along with ambient temperature (black points). (Bottom) Heat input to blocks.}
%     \label{fig:sim-data}
% \end{figure}
%
The top row in Figure \ref{fig:exp-simulated-data} shows results using the proposed method.
% $T_1$ starts heating first, and then shares that heat with $T_2$ which shares it with $T_3$. After roughly 16 minutes, the system reaches equilibrium.
%
The first three subplots (from left-to-right) show the value of $r(T_i, \, T_a = 21)$ for the temperatures that the blocks experienced during the simulation (solid lines) as well as the GPLFM's state estimates (red = $T_1$, blue = $T_2$ and orange = $T_3$; marker size proportional to state variance). The black dotted line with ribbon shows the mean and standard deviation of the posterior predictive distribution in Eq.~\ref{eq:post-preds}, under a third-order Bayesian regression model with $\mu_{i0} = [0 \, 0 \, 0]^{\intercal}, \Sigma_{i0} = 10^3 I$. To test this estimate of the convection function, we simulated the system forward (Eq.~\ref{eq:heatdynamics}) for $1000$ seconds under starting temperatures of $25 \, \degree \mathrm{C}$ and with heat input of 100 $\mathrm{W}$ from $120$ to $600$ seconds. The fourth subplot shows the true simulation (dots) and a simulation using the mean of the posterior predictive (i.e., Eq.\ref{eq:pp_mean}; solid lines). The fifth subplot shows the absolute error between the simulations, producing an overall root mean squared error over time of $0.023$.

We compare the GPLFM with an offline estimate of the effect of the nonlinear convection function. First, we run the state-space model without a nonlinear convection term on the simulated temperature measurements (note that this requires a small noise injection, here $Q = 10^{-12}I$). Then, we calculate the post-fit residuals, i.e., $y_{ik} - m_{ik}$ for all blocks $i$ and time $k$, and fit a standard third-order polynomial regression model on the residuals. The first three subplots on the bottom row of Figure \ref{fig:exp-simulated-data} show the residuals as a function of the temperature state estimates (red = block 1, blue = block 2, orange = block 3), with the estimated polynomials overlaid (black dotted line) which fit well. 
% Note that the polynomials fit well. 
The fourth subplot shows the effect of simulating the temperatures forward (Eq.~\ref{eq:heatdynamics}) with the polynomial regression function instead of the true convection function, and the fifth subplot shows the absolute error between the simulation with the true convection function and the regression function. The root mean squared error over time is $1.385$, which is much larger than that of GPLFM and highlights the importance of estimating the effect of nonlinear convection online.

\begin{figure}[htb]
    \centering
    \includegraphics[width=\columnwidth]{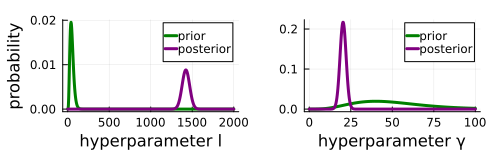}
    \caption{Prior and approximate posterior distributions under Laplace's method for the kernel hyperparameters $l$ and $\gamma$.}
    \label{fig:Laplace-hparams}
\end{figure}

For the hyperparameter optimization of the GPLFM, the prior distributions had parameters $\alpha_l = \alpha_\gamma = 5.0$ and $\beta_l = \beta_\gamma = 0.1$. 
% These favour estimates in the region of $1$ to $100$. 
Figure \ref{fig:Laplace-hparams} show the prior and posterior distributions for each parameter. For $l$, the posterior distribution moves to much larger length scales but also becomes wider indicating that there is a high degree of uncertainty on this estimate. For $\gamma$, the posterior concentrates sharply around roughly $20$, with high certainty.

% \begin{figure}[ht]
%     \centering
%     \includegraphics[width=\columnwidth]{figures/SSM+GPr-block1_fnest.png} \vspace{-20pt} \\
%     \includegraphics[width=\columnwidth]{figures/SSM+GPr-block2_fnest.png} \vspace{-20pt} \\
%     \includegraphics[width=\columnwidth]{figures/SSM+GPr-block3_fnest.png}
%     \caption{Nonlinear convection effects estimated by standard Gaussian process fit to residuals. Estimates are worse than for the Gaussian process augmented state-space model.}
%     \label{fig:sim-GPfits-baseline}
% \end{figure}

% \begin{figure}
%     % \centering
%     \includegraphics[width=\columnwidth]{figures/NONLCONV_SSM+GPr-simulations.png}
%     \caption{Simulation using identified nonlinear convection function (3-order polynomial) found by standard GP fit to white-box model residuals. }
%     \label{fig:sim-predictions-baseline}
% \end{figure}

\begin{figure*}[thb]
    % \centering
    \includegraphics[width=.31\textwidth]{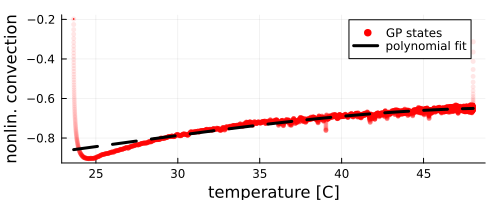} \hfill
    \includegraphics[width=.31\textwidth]{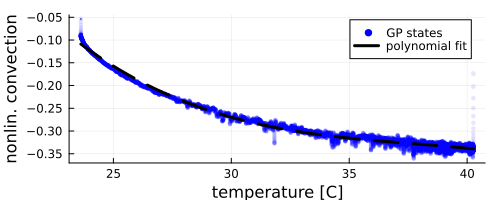} \hfill
    \includegraphics[width=.31\textwidth]{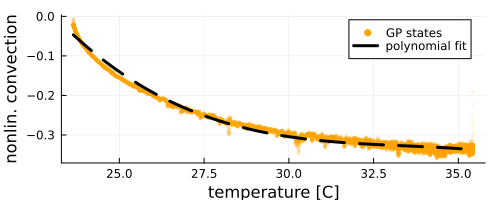} \\
    \includegraphics[width=.31\textwidth]{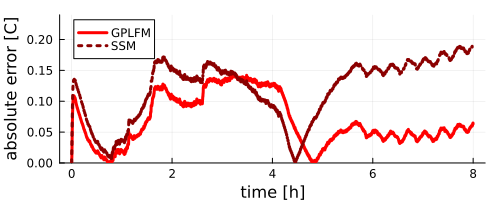} \hfill
    \includegraphics[width=.31\textwidth]{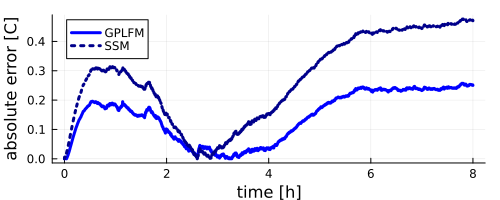} \hfill
    \includegraphics[width=.31\textwidth]{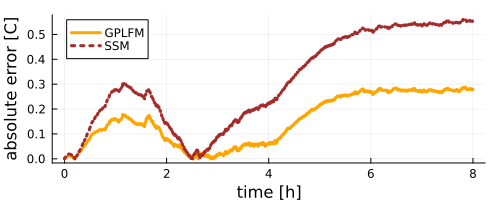}
    \caption{(Top row) GP states as a function of temperature states for sensors $4$ (red), $9$ (blue), and $12$ (orange), with 3d-order polynomial fits (mean posterior predictive distributions; black dashed). (Bottom row) Absolute simulation error,  $|$ measurement - mean prediction $|$, for GPLFM versus SSM method.}
    \label{fig:mdata-GPfits}
    \vspace{-5pt}
\end{figure*}

\subsection{Measured data} \label{sec:measureddata}
In the physical device, block 1 weighs $m_1 = 0.276$ kg, block 2 weighs $m_2 = 0.160$ kg and block 3 weighs $m_3 = 0.74$ kg. The specific heat capacity of aluminum is $c_{ip} = 910 \, \mathrm{J}/(\mathrm{kg} \mathrm{K})$. There are 13 thermistors, with \#1 to \#5 on block 1, \#6 to \#10 on block 2 (sensor \#7 malfunctioned), and \#11 to \#13 on block 3. The measurements within a block are highly similar, and  we thus consider the middle ones as representative of the three components (\#4 = block 1, \#9 = block 2, and \#12 = block 3). The conductance parameters of the pads as well as the heat transfer coefficient $h_a$ are identified during a separate experiment after reaching a steady-state, yielding $k_{12} = 0.272 \, \mathrm{W}/(\mathrm{m} \mathrm{K})$,  $k_{23} = 0.218 \, \mathrm{W}/(\mathrm{m} \mathrm{K})$ and $h_a = 7.75 \, \mathrm{W}/(\mathrm{m}^2 \mathrm{K})$. The surface areas of the blocks are $a_1 = 0.0066 \, \mathrm{m}^2$, $a_2 = 0.0055 \, \mathrm{m}^2$ and $a_3 = 0.0037 \, \mathrm{m}^2$. 

Figure \ref{fig:mdata-GPfits} (top row) shows the GP states (marker size proportional to state variance) as a function of the temperature states (block 1 = red, block 2 = blue, block 3 = orange), with the mean of the posterior predictive distribution of the polynomial regression model overlaid (dotted black line). In general, the GP states are fairly smooth as a function of temperature and the polynomial fits well. Only the first block shows something which is not picked up by the polynomial regression model: a rapid drop when the system first starts deviating from the ambient temperature. The bottom row in the figure compares the open-loop simulation of the state-space model \emph{without} an identified nonlinear convection function (SSM) versus \emph{with} the mean posterior predictive of the polynomial regression (Eq. 41; GPLFM). The error is the absolute difference between the measured temperature data and the simulated temperatures (lower is better). Note that the error of GPLFM is smaller than SSM nearly everywhere for all three blocks, with block 1 showing the smallest difference between methods.
% The ambient temperature does not vary much over the course of the trial, so we plot the nonlinear convection estimate as a function of the block temperature.

\section{Discussion} \label{sec:discussion}
The value of identifying the nonlinear convection function quickly and with quantified uncertainty lies in that it enables optimizing a subsequent step in the control process, such as correcting the positioning signal or placing cooling fans.

We chose the exponential kernel covariance function as it generates a scalar SDE. One could alternatively choose a higher-order Whittle-Mat{\'e}rn, but that leads to vector SDE's. For example, a smoothness parameter of $3/2$ generates a two-dimensional system and $5/2$ generates a three-dimensional one \cite{hartikainen2010kalman}. Recall that the smoothing procedure scales cubically in state-space dimensionality, $O(D^3 N)$. $D$ is related to the dimensionality of the SDE $L$ through $D = (2 + L)M$. Increasing $L$ increases the computational cost drastically. 

% Laplace's method for approximating the posterior distribution of the kernel hyperparameters can be extended to variational Laplace, where the posterior uncertainty also affects the state estimates \cite{daunizeau2017variational}.

% In reality, the nonlinear convection will be a function of the ambient temperature $T_a$ as well. It remains an open question as how to incorporate this dependency in the augmented state-space model.

\section{Conclusions}
We proposed a procedure to estimate the effects of nonlinear convection in a lumped-element heat transfer dynamics model. The procedure is based on a state-space model with known conductance and linear convection effects, augmented with a Gaussian process. Through Bayesian smoothing, we obtain states for temperatures and through Laplace's method, we obtain approximate posterior distributions for the GP kernel hyperparameters. We fitted a Bayesian polynomial regression model that predicts the GP states from the temperature states and ambient temperature measurements, and used it to simulate the effects of nonlinear convection.

\addtolength{\textheight}{-12cm}   % This command serves to balance the column lengths

%%%%%%%%%%%%%%%%%%%%%%%%%%%%%%%%%%%%%%%%%%%%%%%%%%%%%%%%%%%%%%%%%%%%%%%%%%%%%%%%


%%%%%%%%%%%%%%%%%%%%%%%%%%%%%%%%%%%%%%%%%%%%%%%%%%%%%%%%%%%%%%%%%%%%%%%%%%%%%%%%

\bibliographystyle{ieeetr}
\bibliography{references}

\end{document}